\documentclass[acmtog,authorversion,nonacm]{acmart}
\AtBeginDocument{%
  \providecommand\BibTeX{{%
    \normalfont B\kern-0.5em{\scshape i\kern-0.25em b}\kern-0.8em\TeX}}}

\setcopyright{acmcopyright}
\copyrightyear{2023}
\acmYear{2023}
\acmDOI{XXXXXXX.XXXXXXX}

\setcopyright{acmlicensed}\acmConference[ASSETS '23]{The 25th International ACM SIGACCESS Conference on Computers and Accessibility}{ October 22–25, 2023}{ New York, NY, USA}
\acmBooktitle{The 25th International ACM SIGACCESS Conference on Computers and Accessibility (ASSETS '23, October 22–25, 2023, New York, NY, USA}

\begin{document}

\title{Ethical Concerns when Working with Mixed-Ability Groups of Children}

\author{Patricia Piedade}
\affiliation{%
  \institution{ITI, LARSYS, Instituto Superior Técnico, University of Lisbon}
   \city{Lisbon}
\country{Portugal}}
\email{patricia.piedade@tecnico.ulisboa.pt}

\author{Ana Henriques}
\affiliation{%
  \institution{ITI, LARSYS, Instituto Superior Técnico, University of Lisbon}
   \city{Lisbon}
\country{Portugal}}
\email{ana.gfo.henriques@campus.ul.pt}

\author{Filipa Rocha}
\affiliation{%
   \institution{LASIGE, Faculdade de Ciências, University of Lisbon}
   \city{Lisbon}
\country{Portugal}}
\email{fgrocha@fc.ul.pt}

\author{Isabel Neto}
\affiliation{%
   \institution{INESC-ID, Instituto Superior Técnico, University of Lisbon}
   \city{Lisbon}
\country{Portugal}}
\email{TBD}

\author{Hugo Nicolau}
\affiliation{%
   \institution{ITI, LARSYS, Instituto Superior Técnico, University of Lisbon}
   \city{Lisbon}
\country{Portugal}}
\email{TBD}


\begin{abstract}
Accessibility research has gained traction, yet ethical gaps persist in the inclusion of individuals with disabilities, especially children. Inclusive research practices are essential to ensure research and design solutions cater to the needs of all individuals, regardless of their abilities. 
Working with children with disabilities in Human-Computer Interaction and Human-Robot Interaction presents a unique set of ethical dilemmas. These young participants often require additional care, support, and accommodations, which can fall off researchers' resources or expertise. The lack of clear guidance on navigating these challenges further aggravates the problem.
To provide a base and address this issue, we adopt a critical reflective approach, evaluating our impact by analyzing two case studies involving children with disabilities in HCI/HRI research.
\end{abstract}




\maketitle

\section{Introduction}

Inspired by and extending Spiel et al.'s work on the micro-ethics of conducting participatory design with marginalized children \cite{Spiel2018}, we present our considerations of the ethics of working with mixed-ability groups of children.

We situate our research within the fields of ethics and inclusive educational technologies, with an expressed concern for empowering marginalized communities (some of which we belong to) to co-create and take an active role in shaping agendas. We engage with these topics with the ultimate goal of moving away from transactional service models and toward more relational ways of thinking and being in the world in order to challenge hegemonic power structures \cite{10.1145/3544549.3582750}.

In that context, we consider it particularly important to include marginalized populations in our work, and, within that, we also highlight the importance of including children as protagonists in participatory research \cite{10.1145/3078072.3079725}. 

With the growth of accessibility research within and as a sub-field of HCI, recent works move towards a more social and relational model where disability is not located within an individual or infrastructure \cite{bennettaccessible, Holloway2019}. Instead, it is enacted through social-material arrangements and practices (i.e., produced through interactions) \cite{kafer2013feminist}.

Specifically, in the case of inclusive educational technologies, research has grown beyond the adaption of materials for individual use by children with disabilities towards the creation of shared solutions that promote group work between children with and without disabilities, allowing them to play and learn together \cite{inclusivestories, ng4, ball}. Participatory and community-led approaches tend to be favored due to their potential to provide future users with agency over the technology developed for them \cite{Holloway2019, inspo, NetoCommunity}. Though this approach has proven highly effective in creating more equitable classroom environments \cite{ngi3, inspo, NetoCommunity}, it is not without its challenges.


Children-centered research comes with its own set of ethical challenges which must be heeded, especially when working alongside marginalized children \cite{Spiel2018}. This is particularly the case in mixed-ability settings \cite{childrendebate}, where the researcher’s standpoint must be observed in the interactions between diverse groups of children with differing understandings of themselves, their peers, and their environments.

This matters because ethics is contingent \cite{10.1145/3544549.3582750}, and our deliberations as researchers are highly dependent upon social contexts and environments \cite{Antle_2017}. Indeed, this is the value of an approach like Komesaroff's micro-ethics \cite{Troubled_Bodies_2020}. Rather than focusing on ineffective sets of predetermined and overarching principles, micro-ethics can zoom in on the smaller scale day-to-day ethical decisions and interactions that occur organically between people.

In educational contexts such as mixed-ability classrooms, micro-ethics encourages educators and students to engage in ethical reflections and decision-making on a more case-by-case basis. It might also prompt researchers, as well as teachers, to consider how their choices, interactions, and pedagogical strategies impact the well-being and development of each child, particularly those who belong to marginalized groups \cite{Spiel2018}.

These considerations, moreover, necessarily imply some level of caregiving, which necessitates that we theorize on our ability to, as researchers, adequately provide it \cite{10.1093/iwc/iww010}. As such, an understanding of care ethics is relevant to any research involving human participants, especially when working with vulnerable populations \cite{Spiel2018}.

Care is an integral part of all human interactions, but it often remains unacknowledged in research reports. Care is, nonetheless, more than theory, it is, fundamentally, practice. Indeed, Joan Tronto identifies in her work four different yet entwined stages of caring \cite{a9050057-47ca-3631-9941-c34425f0d7d3}, from which we draw for our considerations. They are 1) attentiveness: which refers to the inclination to be attentive and aware of the needs of others; 2) responsibility: which involves being willing to take action and respond to meet those needs, showing a sense of duty and care; 3) competence: which relates to the ability to provide effective care, demonstrating skill in addressing the identified needs; and 4) responsiveness: which encompasses considering the perspectives of others as they perceive them and acknowledging the potential for abuse or misuse in the context of caregiving.

Keeping in line with this theoretical background, we detail two separate case studies within our research working with children in mixed-ability settings in order to provide a reflective account of that research and its ethical challenges. We do so with the intention of highlighting the importance of a processual shift to situational ethics that is community-led \cite{Sanders2008} as opposed to the more typical, albeit largely ineffective \cite{Spiel2018}, prescriptive and static models, to collectively build on more viable approaches to ethical deliberations in dynamic contexts.

\section{Mixed-Visual Ability Groups of Children Collaborating in CT activities}
Our first case study recalls the work developed in \cite{10.1145/3544548.3581261}. Children with disabilities are educated in an inclusive approach within mainstream schools demanding new adaptations of support in learning and social activities \cite{10.1145/3434074.3446356}. Computational thinking (CT) is already established in children's educational curriculum. In inclusive education, collaborative coding environments, besides the learning and social benefits \cite{gokhale1995collaborative}, also have the potential to promote inclusive behaviors between people with different abilities. Regarding the recent shift to remote and hybrid collaborative environments, this work discusses the benefits and limitations of remote and co-located collaboration in CT activities among children with mixed-visual abilities.

\subsection{User Study}
The study used a tangible robotic system resembling the Sokoban game \cite{sokoban}. The collaborative CT activities were set up in two environments that varied in presence and proximity between the pair (remote and co-located) with two interdependent roles (one managed the tangible map and robot, while the other programmed the robot's behavior with coding blocks). We conducted within-subjects research to give children the opportunity to solve puzzles in both environments with both roles. A researcher and their Inclusive Education Teacher were always present for each session.

Ten mixed-visual ability dyads between 10 and 17 years old ($M = 12.75 SD = 1.9$) from three inclusive schools in our country participated in the sessions. Through their educators, we asked the children with visual impairments to invite a sighted schoolmate to form pairs. We ensured that all participants were attending 5th-8th grade considering the national curriculum. The participants' legal guardians signed the consent forms, and the children agreed to participate.

All the sessions were video and audio recorded, and we collected data in light of our research question to measure task performance, social behaviors, and user experience.

\subsection{Possible Concerns}
\begin{enumerate}
    \item \textbf{Balance Interference While Preserving Learning Opportunities} - When working with mixed-ability groups of children, we believe it is important to promote an inclusive environment, i.e., where all children feel safe, supported, and free to participate \cite{inclusiveeducation}. When children share a collaborative environment and its tools, it can be challenging for researchers to properly manage the situation without interfering in the research or the children's relationship. In our study, we encountered an illustrative incident of uncooperative behavior between partners when a sighted child took over the coding blocks of his blind partner and finished that puzzle by himself. Neither the researchers nor the teacher intervened during this interaction, as our primary aim was to observe the social dynamics among the children. However, this lack of mutual respect, along with the substitution of agency of the blind child, resulted in an exclusion experience. Regrettably, this exclusion went unnoticed by all parties involved, representing a missed opportunity for a significant learning moment.
    In summary, it is vital to strike a balance between observing natural peer interactions and addressing situations, even after they have occurred, as demonstrated in the example mentioned above. This approach is indispensable for ensuring that the inclusive environment continually offers substantial enrichment for all participants and that valuable learning moments are not wasted.
    
    \item \textbf{Unmet Expectations} - When children are pulled away from routine activities, they build certain expectations. Our study took place during school hours, and children were told they would be playing together with robots and LEGO. It is fair to assume children built high expectations of fun. These circumstances potentially harm the young participants by disappointing them. During our activities, there were moments of congested participation when children had to wait for their partners. The long waiting period promoted moments with no communication (particularly in remote settings) and, therefore, no awareness of the ongoing activity. For instance, in two of the groups, we noticed that some blind children appeared disengaged, with some even lowering their heads onto the table, sleeping, potentially indicating a state of disinterest. To recap, recognizing and managing children's expectations is essential when conducting activities that deviate from their usual routines. Addressing moments of waiting and non-communication is crucial to ensure a more engaging and inclusive experience for all participants, especially in remote settings.
\end{enumerate}

\section{Neurodiverse elementary school classrooms co-designing a robotic game}
Our second case study describes work developed applying the methodological toolkit proposed in a concurrent publication \cite{piedade2023}. This work explored the inclusive potential of co-design methodologies and tangible robotic games within a neurodiverse classroom environment. Though integrated into mainstream schools, neurodivergent (ND) children often face social exclusion from their neurotypical (NT) peers, as the two groups of children often struggle to engage with each other due to different communication styles, preferences, and sensory needs \cite{prop3,morris2023}. Being the minority, ND children often miss out on group play and its fundamental benefits \cite{Fromberg1990,Fromberg1992,garvey1990,huizinga2014homo,Fromberg2012,peds}. HCI games research has done little to address this issue, with most games taking on a medical framework and focusing on single-player solutions for a single diagnosis \cite{prop3}. We aimed to encourage neurodiverse play through the co-designed game and promote classroom inclusion throughout the co-design sessions.

\subsection{Co-Design Sessions}
The co-design sessions pertaining to this project took place over the course of 6 months in a local public elementary school. We engaged with four classrooms (two second grades and two fourth grades), with a total of 81 students (43 girls and 38 boys, 6-12 years $M=8.22$ $SD=1.26$, 19 ND: 13 learning differences, one dyslexia, two intellectual disabilities, two ADHD, one Down's Syndrome, and one Global Developmental Delay). 

Our process was broken down into five 90-minute sessions encompassing multiple methods (e.g., crafting activities, Expanded Proxy Design \cite{prop1}, low-fidelity prototyping). The first two sessions aimed to familiarise the children with the robotic element they were to work with, a commercial Ozobot \cite{ozobot_2022} robot. The last three sessions focused on the development of game prototypes.

Prior to the co-design sessions, we held a focus group with educators of neurodiverse classrooms and multiple interviews with neurodivergent adults to inform us of the challenges and opportunities we might encounter in the classroom. The children's legal guardians and the participating teachers signed the consent forms, and the children agreed to participate. All the sessions were video and audio recorded, and we collected data in light of our research question to analyze social behaviors and user experience.

\subsection{Possible Concerns}

\begin{enumerate}
    \item \textbf{Transparency vs. Exposure} - When working with a vulnerable population such as children, especially in the case of marginalized children, we believe it is important to communicate our research goals and outcomes clearly. However, with neurodivergence being somewhat invisible, mentioning it within the classroom could bring undue attention to neurodivergent students, which could lead to further ostracization. We elected not to communicate this facet of our research to the children, simply stating, "We are going to create a game everyone in the classroom can play". We utilized techniques, like Expanded Proxy Design \cite{prop1}, to emphasize the needs of neurodivergent children without spotlighting their differences. This method proved effective in making NT children aware of said needs, and one girl with an intellectual disability openly and joyfully stated that the proxy was like her. Nevertheless, this impacted how the design process was conducted, not allowing full transparency with our co-designers. 
    \item \textbf{Teachers' Influence} - As the authority figure within the classroom, teachers hold major sway in any interactions that happen within it. From our initial educator focus group, we understood that they saw themselves as problem solvers. However, ND adults warned us that a teacher's treatment of ND children, be it good or bad, will influence how the NT children treat their ND classmates. Our time in the classrooms validated these concerns and showed us the impact of different teaching styles on neurodiverse group dynamics. In one of the classrooms, a very caring teacher often acted in a coddling way towards her ND students. This was mirrored by NT classmates, who did not exclude ND students but didn't see them as equals either. In another classroom, an assertive teacher often solved group conflicts by demanding everyone perform the task in the same \textit{neurotypical} way, barring creative freedom and undermining neurodivergent interpretations. In both cases, we recognized an issue but did not feel comfortable intervening given the existing hierarchy, which may have been a choice in detriment of the participating children. It is essential to highlight that none of the teachers acted in bad faith.
    \item \textbf{Balancing Opinions} -  As a direct result of us not communicating the ND aspect of our study, all group members (NT and ND) were seen as equal. This posed a problem when it came to group decision-making. Children often struggled to find a single solution that would fit all of their needs and preferences. When this happened, they tended to use voting as decision-making. Within this scenario, the fact that NT children were the majority put ND interests and needs at a bigger risk of being ignored. To circumvent this issue, we tried to work with the groups towards compromising on ideas that mixed multiple ideas rather than choosing a single one. Nevertheless, it is unclear how to make ND voices heard within these group contexts without bringing undue exposure. Though direct mediation proved somewhat effective in our case, the presence of a researcher during this creative activity may have also stunted the full creative potential of child-led ideation.
    \item \textbf{Classroom Expectations} As pointed out by Spiel \& Gerling in their review of HCI games research with ND populations \cite{prop3}, classroom environments are not the most hospitable for ND self-determination. Working within them is, nevertheless, important as children spend a significant amount of time in these environments. The typical classroom rules (e.g., sitting still, being quiet) are unnecessary for co-design activities and may even be counterproductive in many cases. However, with the limited space and acoustics, some classroom management is needed to maintain a sustainable environment for all participants. On several occasions, we witnessed ND children, primarily one boy with ADHD, being scolded by both teacher and classmates for behaviors such as stimming, frequently getting up, and getting off-task. As researchers, we were aware such behaviors are to be expected and healthy, and we wanted to encourage them. However, our perception limited the authority within the classroom and stopped us from changing this status quo in favor of a safer, more inclusive working environment.
\end{enumerate}

\section{Conclusion / Future Work}

As we can see in the preceding case studies, working with children in a mixed-ability setting comes with several added responsibilities and ethical concerns \cite{Holland_Renold_Ross_Hillman_2010,childrendebate}, which illustrates the need for a more robust approach to dealing with such complexities. This underscores the necessity for a more comprehensive approach to address these complexities, particularly in terms of researchers' and teachers' involvement in children's peer interactions, the appreciation of individual differences without stigmatization, and the continuous effort to maintain engaging and accessible activities that align with the participants' expectations.

Faced with these challenges, we recognize the benefits of a participatory approach to our research toward ethics \cite{10.1145/3544549.3582750} and inclusive educational technologies \cite{10.1145/3544548.3581261,piedade2023}. We are, however, mindful of the micro-ethics involved in such complex co-design environments \cite{Spiel2018}. To help bridge that gap, we find that an approach rooted in care ethics must help inform these decisions \cite{10.1093/iwc/iww010} through a participatory process to value-sensitive design\cite{Friedman_Hendry_2019}.

Indeed, participatory design, micro-ethics, and care ethics intersect in important ways, especially when working with children in mixed-ability environments. Their intersection points to a more holistic framework for creating inclusive and ethically sound educational environments founded upon ethics that are processual and situational rather than static and prescriptive.

Participatory design emphasizes the active involvement of all stakeholders, including children, in the design and decision-making processes. When applied to mixed-ability settings, this approach ensures that the diverse needs and perspectives of children with varying abilities are considered. Additionally, it empowers these children to have a say in shaping their own learning experiences, thus fostering a sense of agency and inclusion.

Care ethics presupposes that all beings are interconnected and interdependent, highlighting the importance of providing and receiving care as the basis of those interactions \cite{Tronto1990}. In tandem with a participatory approach to research, care ethics brings a more relational understanding of ethics as it occurs in the interstices of the interactions between people — including those between researchers and participants, children and adults, etc. In the context of this work, care ethics highlights the importance of nurturing and sustaining caring relationships within research and educational settings \cite{10.1093/iwc/iww010}. When applied to mixed-ability learning environments, an ethics of care calls for a deep understanding of the unique needs and vulnerabilities of each child, with a focus on fostering a supportive environment that is appropriately conducive for learning, as per Tronto's stages of caring \cite{a9050057-47ca-3631-9941-c34425f0d7d3}. Care ethics thus challenges researchers and educators to prioritize the well-being and emotional development of all children, recognizing that children with disabilities may require care that might deviate from standardized models catering to children who are already mostly likely to thrive under normative settings. 

This last point is especially relevant given the ethos of care ethics, particularly as proposed by Joan Tronto, of increasing the value of counter-hegemonic actions that distribute political power and highlight the importance of the collective \cite{Tronto1993}. In that regard, the goals of both care ethics and participatory design – "aimed at reinforcing democracy by acknowledging and supporting a diversity of voices" \cite{Spiel2018, Halskov_Hansen_2015} – are quite closely aligned. Going even further, however, given the overlap in intentions, we consider community-led design to be a more promising way forward for ethics in HCI and Accessibility. Indeed, community-led design is a movement focused on reframing the approach to co-design with a specific focus on empowering communities to catalyze their own needs through context-based solutions \cite{Sanders2008}.

Beyond those already detailed throughout, there are important challenges to such an approach left to ponder in the future, especially as it relates to working with children specifically. How can we make the shift from prescriptive ethics to situational and processual ethics with the added challenge of centering the personhood of children? How can we ensure that involving parents and teachers as stakeholders does not compromise nor overpower children's autonomy and self-determination in assessing their own needs and values?

\begin{acks}
We thank all the children, their educators, and schools that agreed to participate in our research projects. We also highlight the contributions to the described case studies of each respective paper's co-authors.

This work was supported by the European project DCitizens (GA 101079116), the FCT project UIDB/50009/2020 and scholarships \newline SFRH/BD/06452/2021 and
SFRH/BD/06589/2021, the Portuguese Recovery and Resilience Program (PRR), IAPMEI/ANI/FCT under Agenda C645022399-00000057 (eGamesLab), through the scholarship BL195/2023.
\end{acks}

\bibliographystyle{ACM-Reference-Format}
\bibliography{sample-base}


\end{document}